# IMPACT OF ROAD MITIGATION MEASURES ON AMPHIBIAN POPULATIONS: A STAGE-CLASS POPULATION MATHEMATICAL MODEL

**Renaud Jolivet, Michel Antoniazza, Catherine Strehler-Perrin and Antoine Gander**

Groupe d'Etude et de Gestion Grande Cariçaie, Champ-Pittet, CH-1400 Yverdon-les-Bains, Switzerland

*CONTACT INFORMATION*
e–mail: renaud.jolivet@a3.epfl.ch

**Abstract**

It is now well established that amphibians are suffering widespread decline and extinctions. Among other causes, urbanization is responsible for habitat reduction, habitat fragmentation and massive road kills. In this context, it is urgent to develop and assess appropriate conservation measures. Using yearly censuses of migrating adults of two anuran species at one location in Switzerland, we examined the impact of a road mitigation measure – permanent under-road tunnels with guiding trenches – along a road separating wintering forests from breeding wetlands. We observe that the adult migrating populations do not exhibit any long-term trend but undergo a transient increase a few years after the installation of the road mitigation measure. Using additional datasets like climatic data and censuses obtained in a control area, we show that the observed pattern of migrating populations cannot be explained by any other data at our disposal. We then checked as a working hypothesis whether the installation of under-road tunnels could explain the observed transient or not. To this end, we use a simple population model and show that the road mitigation measure together with competition for resources can successfully explain the experimental observations. We conclude by discussing the requirements for further assessment of this hypothesis as well as consequences for conservation planners.

# INTRODUCTION

After years of controversy [1], it is today clearly established that amphibians are suffering widespread decline and extinctions [2-4]. Several causes for this decline have been proposed. A non-exhaustive list includes climatic changes, exposure to ultraviolet-B and increased prevalence of diseases [1,2,5,6]. Apart from these, urbanization and roads have been shown to negatively impact amphibian populations [7-11] as well as other taxa [12] and provoke a global degradation of ecological conditions [13]. They are responsible for habitat reduction, habitat fragmentation [8,14,15] and massive road kills [8,16-19], essentially during seasonal migrations between wintering and breeding sites.

While it is difficult to fight against climatic changes and diseases, it should be relatively easy to reduce the negative impact of roads by constructing permanent under-road tunnels with guiding trenches along the main migration corridors. In principle, these tunnels should allow the amphibians to cross the road without danger and should partially restore the connectivity between natural milieus, thus decreasing insularity. Although quite expensive (construction alone reaches about 300000 €/km for the system monitored in this paper), this type of measure has been widely implemented in Europe [20]. However, the question remains open whether it is cost-effective in terms of population conservation or not. In a context of ever-increasing human land use [13], increased importance of land conservation [21] and declining amphibian populations in highly urbanized regions [3], this is a crucially important question to answer.

Here, we report about a long-term survey of two amphibian populations at a single site in Switzerland where permanent under-road tunnels were installed in migration corridors. Urbanization is a particularly critical issue on the Swiss plateau (located between the Alps and the Jura mountains) which is one of the most urbanized regions of the world [~430 inhabitants/km$^2$ [22]]. There, amphibian habitats have been under constant pressure during the last decades. Numerous wetlands and areas prone to flooding have been drained from the early 19th century and forests have been cleared to accommodate the growing need for new land. Concomitantly, the landscape is fragmented by a very dense network of roads and railways. During the 1980s, conservation planners put government agencies under pressure to install road mitigation measures in important migration corridors [23]. Following this line, three sectors of tunnels with guiding trenches were constructed in 1991 in one area known as Cheseaux in the Grande Cariçaie natural reserve under a national road which separates breeding wetlands from wintering forests (supplementary Figure 1). In most amphibian studies, the breeding and wintering sites constitute a highly patchy environment causing problems in data interpretation [1]. In contrast to this, the Grande Cariçaie reserve is an essentially linear structure. Wetlands extend along the south-eastern bank of the Lake of Neuchâtel. Immediately to the east, forests provide wintering habitats for the amphibians (supplementary Figures 2 and 3). This reserve is therefore ideally suited to study the impact of the road mitigation measure.

In order to assess the efficacy of amphibian tunnels as a conservation measure, we censused the populations of yearly migrating adults of common toad (*Bufo bufo*) and common frog (*Rana temporaria*) before and after the construction of the road mitigation system operational for the first time in 1992 in the Cheseaux area (supplementary Figure 2). Starting in 1994, the same species were also censused in a second area known as Ostende where no road passes inside the natural reserve to serve as control. The Ostende area is about 30 km distant from Cheseaux. It is structurally very similar to the Cheseaux area except that it is free of roads. In particular, breeding and wintering sites follow the same spatial organization than in Cheseaux (supplementary Figure 3). We can then use populations censused in this area as a control for some conditions affecting the success of reproduction like the level of the lake that regularly floods breeding sites as well as some climatic data. Other amphibian species have been observed in large numbers in the reserve like pool frog (*Rana esculenta*), alpine newt (*Triturus alpestris*) and smooth newt (*Triturus vulgaris*). However, only common toad and common frog have significant populations in the area where the tunnels were built, preventing us to make comparisons for the former species. The population of both common toad and common frog can be considered stable in both areas. Interestingly, both species have a similar breeding ecology [24].

In this study, we observed that the adult populations censused in both the Cheseaux area (with mitigation measures) and in the Ostende area (without road) do not exhibit any long-term trend. However, both populations censused in the area equipped with under-road tunnels and guiding trenches exhibit a very significant transient increase four years after the installation of tunnels. The paper is organized as follows. We first report and discuss the results of the census. In a second step, we show that the observed transient increase cannot be correlated with any other supplementary dataset at our disposal. Assuming as a working hypothesis that the installation of tunnels could be the cause for the observed transient, we then show using a simple population model that the road mitigation measure and the competition for resources could collaboratively explain the experimental observations. Finally, we discuss the requirements for further assessment of this hypothesis and its potential implications for conservation planners.

# MATERIALS AND METHODS

## CENSUS

In the Cheseaux area (with mitigation measures), populations were censused once in 1983 and then every year between 1992 and 2004. The under-road tunnels and guiding trenches were operational for the first time in 1992. In this area, we captured adult amphibians during the spring migration with bow-nets installed at tunnel exit on the west side of the road except in 1983 when we used drift fences with pit traps instead. The numbers of captured adults in this area were in average 680 (common frogs) and 320 (common toads). In the Ostende area, adult populations were censused yearly during the spring migration between 1994 and 2004. We used drift fences with pit traps. The drift fences were placed so that they were roughly located at the same structural level in the environment than the road in Cheseaux, i.e. just before breeding sites. The numbers of captured adults in the Ostende area were in average 310 (common frogs) and 85 (common toads). In both areas, the setup was designed so as to census the population of migrating adults homing to the breeding sites. The whole installation (bow-nets, drift fences and traps) was set up every year in late January. Every morning, when the conditions were favourable to amphibian migration [rainy night and temperature approximately higher than 4–5°C at dusk [24]], we emptied traps and bow-nets and freed amphibians after specie, gender and approximate age determination by measuring the size of the body. The census was stopped few weeks after the return migration of adults was reported, usually in late April. Returning juveniles were censused in the Cheseaux area once in 1991 (July-August, same collection and determination protocol than for adults) and once in 1992 (June-July). Juveniles were captured with pit-traps on the eastern side of the road. A map of the region and aerial views of Cheseaux and Ostende areas can be seen in supplementary Figures 2 and 3.

## ESTIMATION OF TRAFFIC LOAD

The daily traffic distribution on the road protected with amphibian tunnels was extrapolated from the Swiss Standard SN 640 005a [25]. The average daily traffic load on this road is estimated to be 5500 vehicles/day (data provided by the Service des routes de l'Etat de Vaud, yearly average). We assume that it is composed of half *local traffic* and half *leisure traffic* (2750 vehicles/day each). Time series in Figure 3a were computed (weighted sum) using the seasonal variations of these two types of traffic. The local traffic intensity is at 101.5% of the yearly average in March-April and at 104% in July. Values for the leisure traffic intensity are 94% for March-April and 131% for July. Our extrapolation for March-April is in excellent agreement with direct counts performed between April 12$^{th}$ 2005 and April 19$^{th}$ 2005 (data provided by the Service des routes de l'Etat de Vaud, not shown).

## POPULATION MODEL

Populations are modelled using the Lefkovitch matrix population projection technique [26-28]. The population **N** of each specie at time *t* is given by

$$\mathbf{N}_t = \mathbf{B} \cdot \mathbf{N}_{t-1} \qquad (1)$$

where **N** is a ($T$+1)-dimension vector with $T$ the number of years between hatching and maturation. **N** is composed of the $T$ classes of juveniles $N_0, \ldots, N_{T\text{-}1}$ and of a single class of adults $N_A$. The population projection matrix **B** is given by

$$\mathbf{B} = \begin{bmatrix} 0 & \mathrm{L} & \mathrm{L} & 0 & \beta \\ \alpha_J r_{\lambda_J, t-1} & \mathrm{O} & & \mathrm{M} & 0 \\ 0 & \mathrm{O} & \mathrm{O} & \mathrm{M} & \mathrm{M} \\ \mathrm{M} & \mathrm{O} & \mathrm{O} & 0 & 0 \\ 0 & \mathrm{L} & 0 & \alpha_J r_{\lambda_J, t-1} & \alpha_A r_{\lambda_A, t-1} \end{bmatrix} \quad (2)$$

where $r_{\lambda,t} = \min\left(1, \exp\left[-\lambda\left(N_{M,t} - K\right)\right]\right)$ models the density dependent competition for resources. $K > 0$ can be interpreted as the environment carrying capacity. $\lambda_{A,J} > 0$ is the competition intensity exerted by adults on adults (*A*) and on juveniles (*J*). $0 \le \alpha_{A,J} \le 1$ is the maximal yearly survival rate of adults (respectively juveniles). Finally, $\beta$ summarizes all the steps between the mating of mature individuals at the breeding sites and the return migration of juveniles to the wintering sites. The installation of amphibian tunnels is included in the model by an instantaneous positive shift of $\beta$ in 1992.

### MODEL PARAMETER FITTING

The model was fitted on data by minimization of the least-square distance with a standard algorithm [29]. At each iterations, $\beta$ was first set to the arbitrary value $\beta = 1$ with 100 individuals in each classes and the model was simulated until it reached a quasi-stationary state (typically for 1000 years). $\beta$ was then set to a new value $\beta_{\text{new}} > \beta$ in 1992 to model the positive impact of amphibian tunnels. Finally, the predicted values $\hat{N}_0, \mathrm{K}, \hat{N}_{T-1}$ and $\hat{N}_A$ were rescaled so that the value $\hat{N}_A$ predicted by the model in 1992 corresponds to the observed density. For both species, we found that raw population numbers as predicted by the model, i.e. before rescaling, are in the same range than observed migrating populations (not shown). Fitted parameters are $\alpha_{A,J}$, $\lambda_{A,J}$, $K$ and $\beta_{\text{new}}$. Even though it is assumed that the adult population is not affected by the tunnels, the same quality of predictions can be obtained when a positive shift of $\alpha_A$ is included in 1992 (not shown).

## RESULTS

### POPULATION CENSUSES

In order to assess the efficacy of amphibian tunnels as a conservation measure, we censused the populations of yearly migrating adults of common toad (*Bufo bufo*) and common frog (*Rana temporaria*) in the Cheseaux area before and after the construction of the road mitigation measure operational for the first time in 1992. Starting in 1994, the same species were also censused in a second area known as Ostende where no road passes inside the natural reserve.

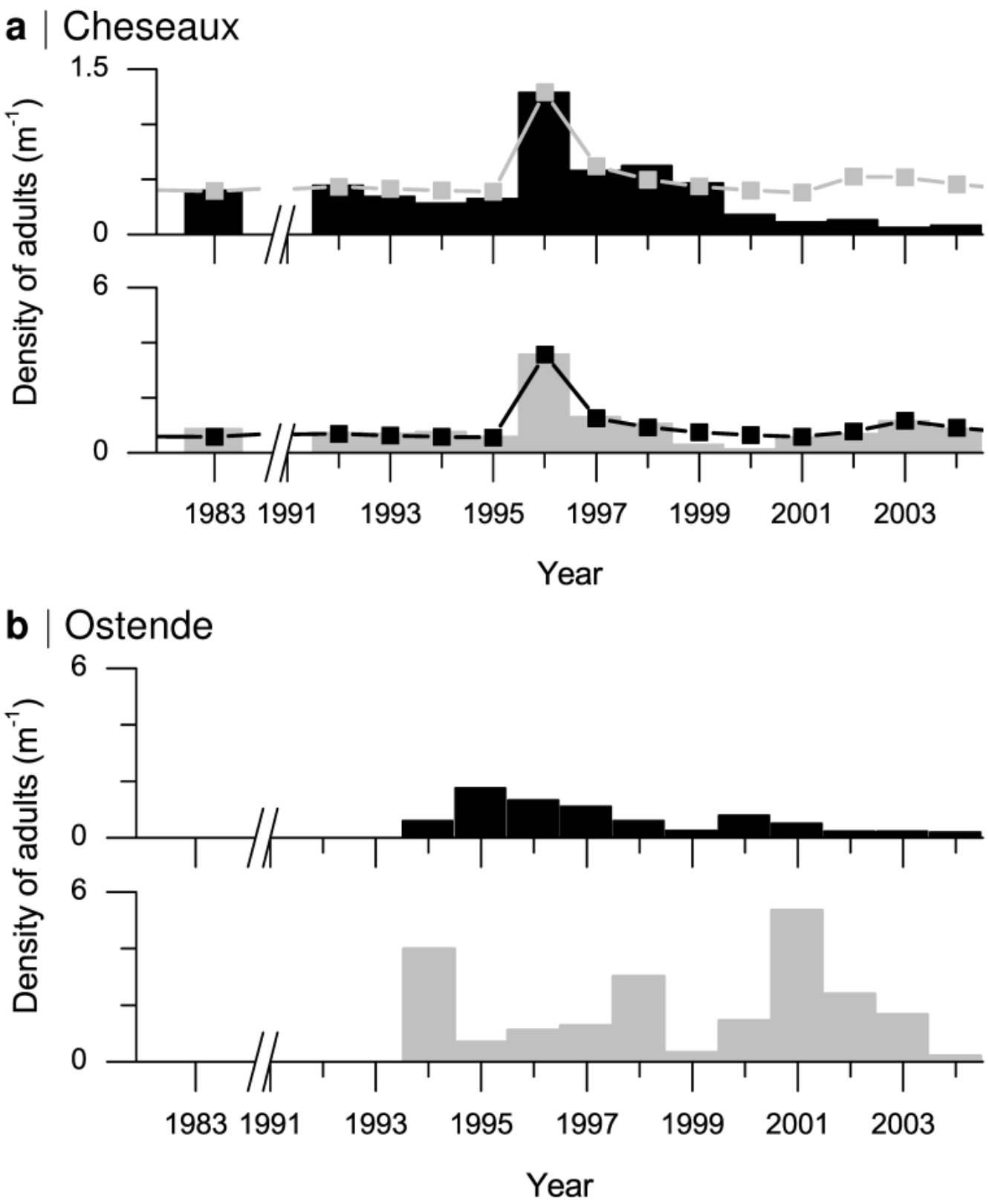

FIGURE 1 RESULTS OF THE CENSUS. (A) DENSITIES OBSERVED IN THE CHESEAUX AREA FOR THE COMMON TOAD (BLACK BARS) AND THE COMMON FROG (GREY BARS; NUMBER OF MIGRATING ADULTS PER METER OF FENCE/TRENCH). THE PREDICTIONS OF THE MODEL ARE PLOTTED FOR BOTH THE COMMON TOAD (BEST LEAST-SQUARE FIT; GREY LINE) AND THE COMMON FROG (BLACK LINE). NOTE THAT COLOURS ARE INVERTED COMPARED TO THE BAR PLOTS. (B) RESULTS OF THE CENSUS IN THE OSTENDE AREA.

Figure 1 shows the results of the censuses in these two areas. We observe that the adult populations censused in both the Cheseaux area (with mitigation measures, Figure 1a) and in the Ostende area (without road, Figure 1b) do not exhibit any long-term significant trend. While the populations in Ostende seem to fluctuate randomly, populations in Cheseaux exhibit a very similar and peculiar pattern with a strong peak in 1996 (adult migrating population is about four times larger than the average level in years 1983-1995 for both species) followed by an almost immediate reset to the precedent level. This type of very strong fluctuations are rare [4]. Analysis of correlation coefficients between the four time series of Figure 1 confirms that the populations of common toads and common frogs in Cheseaux follow a similar dynamics while nothing significant occurs in Ostende (Table 1). We additionally observe that the populations of common frogs are non-declining over the period of the study in agreement with the findings of other authors [30]. The overall population of common toad seem to be in decline after 2000 but long-term data is lacking and so, strong conclusions cannot be drawn [1].

Many causes could lead to the transient increase in migrating populations observed during year 1996. However, it is particularly interesting to note that specie-specific effects can be ruled out since this specific pattern was observed both in common toad and in common frog populations. Similarly, some large-scale variables like the level of the lake that floods the breeding sites can be ruled out since the pattern was observed in Cheseaux but not in Ostende where the same species are present in a very similar environment. In the following, we explore plausible explanations to the migration peak using all data at our disposal that could satisfactorily explain the difference observed between the censuses obtained in Cheseaux and in Ostende, the latter site being used as a control for climatic data.

| Cheseaux | | Ostende | |
| --- | --- | --- | --- |
| common toad | common frog | common toad | common frog |
| 1 | 0.81 ( $P < 5 \cdot 10^{-4}$ ) | 0.53 ( $P = 0.09$ ) | -0.21 ( $P = 0.54$ ) |
| | 1 | 0.37 ( $P = 0.26$ ) | -0.15 ( $P = 0.67$ ) |
| | | 1 | -0.20 ( $P = 0.56$ ) |
| | | | 1 |

TABLE 1 MATRIX FOR THE TIME SERIES CENSUSED IN THE RESERVE (FIGURE 1).

## ASSESSMENT OF POSSIBLE CAUSES FOR THE OBSERVED MIGRATION PATTERN

As we mentioned already, both the Cheseaux and the Ostende areas display the same spatial arrangement of habitats (supplementary Figure 3). They also share a same relation to the lake water level (ground topography is very similar) and no important management measures were applied either in the forests (known wintering sites) or in the wetlands (known breeding sites) during the period of the study. Obstacles used for the census were roughly located at the same structural level in the environment.

Climate can influence the success of the reproduction [6] as well as the migration phenology. While it is usually assumed that temperatures do not vary significantly at this geographical scale (www.meteoswiss.ch), rainfalls could. We thus compared daily rainfalls measured at the Yverdon weather station used as a proxy for the Cheseaux area and at the Payerne weather station used as a proxy for the Ostende area in the years before the peak migration during months corresponding to early stages of life for juveniles [see map in supplementary Figure 2; years 1961-1995; months March-June [24]] and did not found any significant differences (Table 2; paired t-test; data provided by www.meteoswiss.ch). Moreover, the two species have a different egg-laying ecology, suggesting they could be affected in different ways by climatic factors [24]. We also observed no differences in the migration phenology between areas decreasing further the likelihood of a climatic explanation (Figure 2). There is no significant year-to-year trend in the day of the peak migration (linear regression; common frogs in Cheseaux, $R^2 = 0.14$, $P = 0.22$; in Ostende, $R^2 = 0.08$, $P = 0.42$; common toads in Cheseaux, $R^2 = 0.04$, $P = 0.52$; in Ostende, $R^2 = 0.03$, $P = 0.62$). The specie-specific mean is not significantly different between areas (paired t-test; March 2nd for the common frog; $t = -1.11$, $\text{d.o.f.} = 9$, $P = 0.30$; March 11th for the common toad; $t = -0.95$, $\text{d.o.f.} = 9$, $P = 0.37$) and the fluctuations between areas are mildly but significantly correlated (linear fit on pooled data; $R^2 = 0.52$, $P < 5 \cdot 10^{-4}$). These results strongly suggest that the transient increase in migrating populations observed in the area where road mitigation measures are present cannot be satisfactorily explained by climatic data.

Since amphibians are supposed to be structured in metapopulations [1,31,32], the migration peak of 1996 could be explained by strong immigration from a nearby colony. The nearest known group is about 2.5 km away from the northern protected sector and contains both common toads and common frogs. This distance is however rather extreme compared to the typical observed migration distances in this type of environment [33] [but see [32]]. Furthermore, one would expect the immigration to be localized, likely in the nearest sector and spread out in time as migrants disseminate along the bank. We observe on the contrary that the pattern is homogeneously present all along the road except in the population of common toad of the central protected sector which is by far the smallest population (<40 adults in average; supplementary Figure 4). Finally, despite extensive research, we could not find any reason that could explain why representatives of both species would have migrated synchronously in 1996 from this remote site.

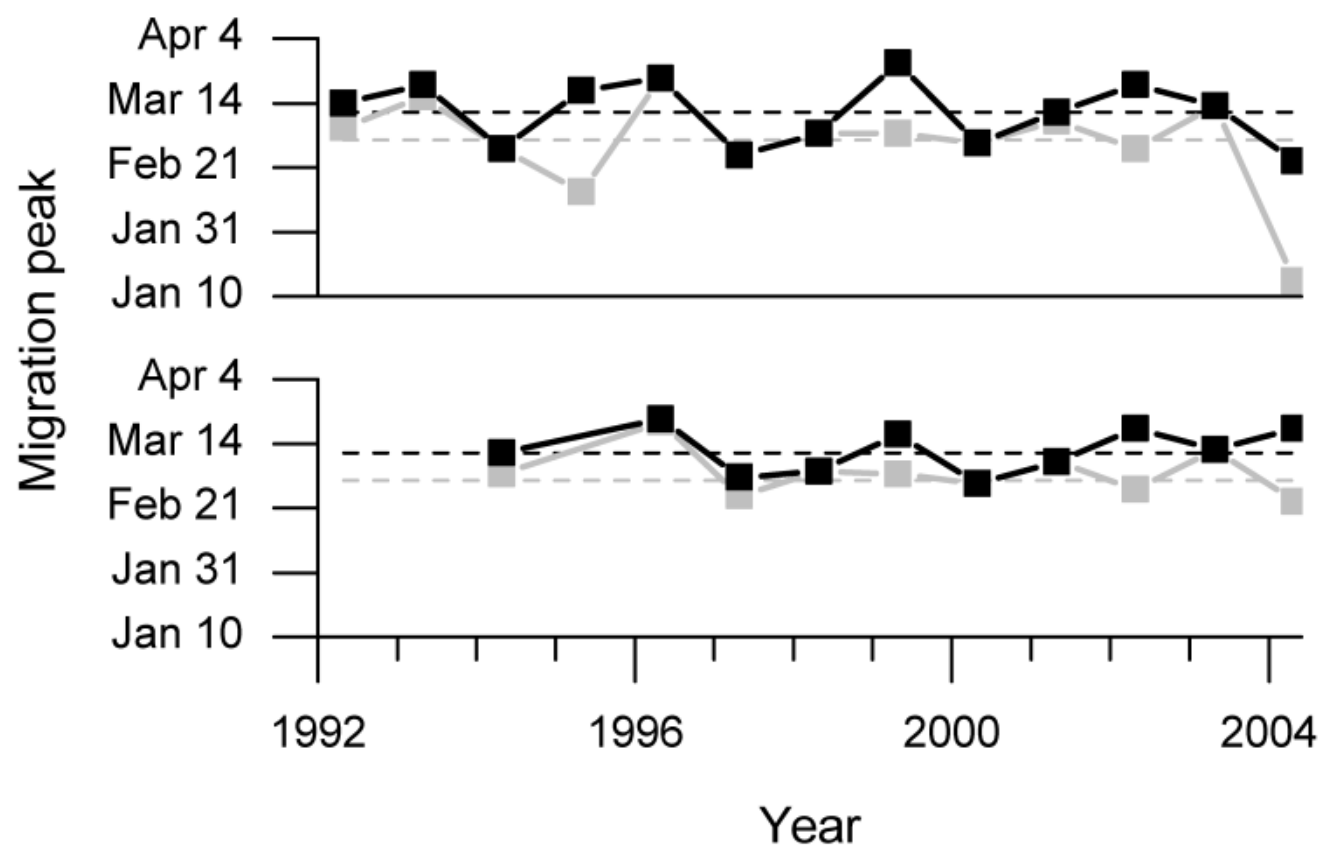

FIGURE 2 MIGRATION PHENOLOGY. PEAK OF THE MIGRATION IN THE CHESEAUX AREA (TOP) AND IN THE OSTENDE AREA (BOTTOM; MISSING PHENOLOGY DATA IN 1995) FOR COMMON TOADS (BLACK) AND COMMON FROGS (GREY). THE MEAN FOR EACH SPECIES IS INDICATED BY A DOTTED LINE.

Interaction with predators may be a potential cause of synchrony in biological populations [see e.g. [34]]. However, the two species do not suffer from predation by exactly the same set of predators. Their main common predators are birds and more specifically in this region, kites and owls. In a survey of bird populations in the reserve, we found that these populations are anecdotic and stable, a few individuals at the maximum (probably <10 in the Cheseaux area).

In absence of any other readily available explanation, it is then tempting to check whether the installation of under-road tunnels could explain the observed transient increase or not. Of course, the specific pattern observed in the Cheseaux area does correlate with the installation of the under-road tunnels and guiding trenches. It is not clear however why the effect is delayed by four years and short-lasting. In the next section, we propose a simple stage-class population model including the effect of the road mitigation measure plus competition for resources that satisfactorily accounts for experimental observations.

## IMPACT OF THE ROAD MITIGATION MEASURE IN A POPULATION MODEL

We constructed a population model based on the Lefkovitch matrix population projection technique [26-28]. For the sake of simplicity, the model divides the population in five classes, one class per year in the juvenile stage [four for each specie in these populations; G. Berthoud, unpublished observations, assessed by skeletochronology; see also [35]] and one class of adults. We further assume that all the adults participate to the migration and do not distinguish between males and females. The model includes density dependent competition which has proven a key component in population models [36]. The competition is parameterized by one parameter describing the carrying capacity of the environment ($K$) and is exerted by adults on adults and juveniles. The mitigation measure is included in the model by an instantaneous positive shift in 1992 of the parameter that describes the overall success of the reproduction ($\beta$) but we make the simplifying assumption that they do not affect the survival rate of adults ($\alpha_A$). This is motivated as follows. The prenuptial migration of adults across the road usually takes place during a few favourable nights in March and they return to the forest after breeding in April. Adults migrate approximately between 7 p.m. and 4 a.m. [19]. Juveniles usually emerge from the breeding sites in late June and migrate *en masse* often during the day. The diurnal traffic load is larger than the nocturnal one and similarly, the summer traffic load is larger than the spring traffic load due to tourism (Figure 3a). We therefore expect the juveniles to be more sensitive to the installation of tunnels than adults. Confirming this hypothesis, the density of migrants was not significantly affected by the tunnels on the short-term (Figure 1a; 1983-1995) even though the road-kills during the spring migration were significantly reduced to almost zero level (not shown). On the other hand, a comparison between censuses of juveniles during summer 1991 and 1992 has shown a sharp five-fold increase of the migration density consecutive to the installation of tunnels for both species (juveniles censused with two pit traps; 585 common toads and 56 common frogs in 1991; 3114 common toads and 273 common frogs in 1992).

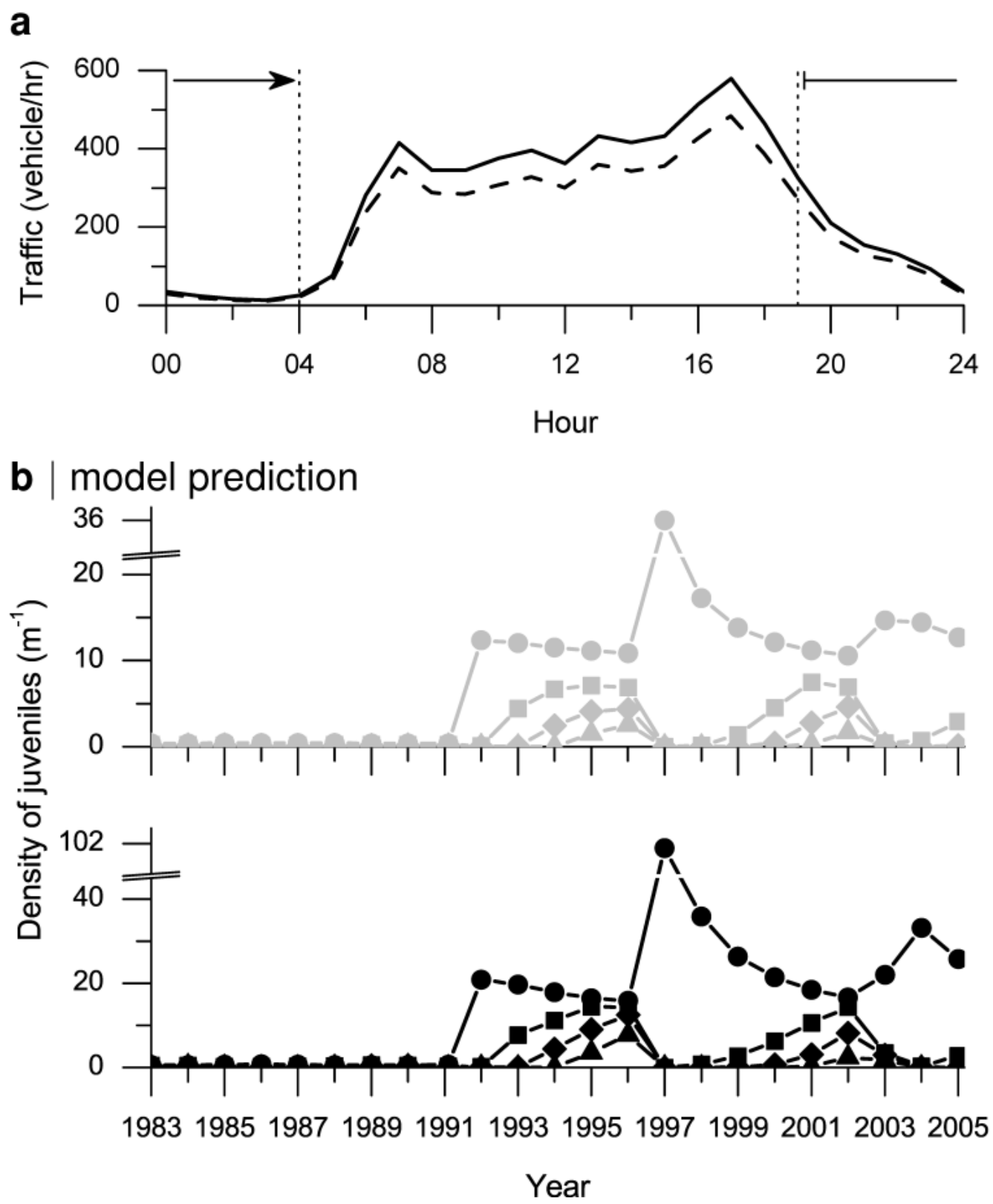

FIGURE 3 TRAFFIC AND MODELLED IMPACT ON JUVENILES. (A) ESTIMATED DAILY TRAFFIC DISTRIBUTION ON THE ROAD PROTECTED BY TUNNELS IN MARCH-APRIL (DASHED LINE) AND IN JULY (SOLID LINE). THE ARROW INDICATES THE BEGINNING (7 P.M.) AND ENDING (4 A.M.) OF THE MIGRATION OF ADULTS. (B) DENSITY OF THE DIFFERENT GENERATIONS OF JUVENILES AS PREDICTED BY THE MODEL FOR THE COMMON TOAD (GREY LINE) AND THE COMMON FROG (BLACK LINE; ● 1ST, ♦ 2ND, ■ 3RD AND ▲ 4TH GENERATIONS). NOTE THAT COLOURS ARE INVERTED COMPARED TO THE BAR PLOTS IN FIGURE 1A.

| Year | $t$ | $P$ | $\|\Delta\mu\|$ (mm/day) | Year | $t$ | $P$ | $\|\Delta\mu\|$ (mm/day) |
|---|---|---|---|---|---|---|---|
| 1961 | 1.25 | 0.21 | 0.94 | 1979 | 0.24 | 0.81 | 0.06 |
| 1962 | 0.57 | 0.57 | 0.18 | 1980 | 0.11 | 0.91 | 0.04 |
| 1963 | 0.18 | 0.86 | 0.07 | 1981 | 2.35 | 0.02 | 0.39 |
| 1964 | -1.61 | 0.11 | 0.51 | 1982 | -0.31 | 0.76 | 0.06 |
| 1965 | 1.19 | 0.24 | 0.32 | 1983 | 0.32 | 0.75 | 0.08 |
| 1966 | -1.56 | 0.12 | 0.25 | 1984 | -0.30 | 0.77 | 0.06 |
| 1967 | -0.59 | 0.56 | 0.17 | 1985 | 1.00 | 0.32 | 0.24 |
| 1968 | -0.94 | 0.35 | 0.19 | 1986 | -1.38 | 0.17 | 0.45 |
| 1969 | 0.45 | 0.65 | 0.12 | 1987 | -1.20 | 0.23 | 0.36 |
| 1970 | -0.25 | 0.80 | 0.08 | 1988 | 0.84 | 0.40 | 0.36 |
| 1971 | -0.66 | 0.51 | 0.25 | 1989 | -1.02 | 0.31 | 0.33 |
| 1972 | -1.73 | 0.09 | 0.44 | 1990 | -0.93 | 0.35 | 0.37 |
| 1973 | -2.39 | 0.02 | 0.81 | 1991 | 0.45 | 0.65 | 0.10 |
| 1974 | 0.62 | 0.54 | 0.11 | 1992 | -1.75 | 0.08 | 0.28 |
| 1975 | -1.62 | 0.11 | 0.30 | 1993 | -1.02 | 0.31 | 0.29 |
| 1976 | -2.16 | 0.03 | 0.37 | 1994 | -0.52 | 0.60 | 0.20 |
| 1977 | -0.89 | 0.37 | 0.32 | 1995 | -1.07 | 0.28 | 0.23 |
| 1978 | 0.03 | 0.98 | 0.01 | | | | |

TABLE 2 COMPARISON OF RAINFALLS BETWEEN OSTENDE AND CHESEAUX AREAS. PAIRED T-TEST (MARCH-JUNE); D.O.F.=121 (STANDS FOR DEGREES OF FREEDOM); $|\Delta\mu|$ STANDS FOR THE ABSOLUTE VALUE OF THE DIFFERENCE OF THE MEANS.

In order to check for the ability of the model to account for observed data, the modelled adult population was fitted on census data of the Cheseaux area. Figure 1a shows that the model successfully explains the observed pattern of migration for both common toads and common frogs. The correlation coefficient measured between modelled and observed data is highly significant (best least-square fit; $R^2 = 0.70$, $P < 4 \cdot 10^{-4}$ for the common toad and $R^2 = 0.94$, $P < 10^{-4}$ for the common frog). Best fit parameters are $\beta_{\text{new}} = 27.89$, $\alpha_J = 0.62$, $\alpha_A = 0.94$, $1/\lambda_J = 62.5$, $1/\lambda_A = 2000$, $K = 599$ for common toads and $\beta_{\text{new}} = 28.52$, $\alpha_J = 0.87$, $\alpha_A = 0.92$, $1/\lambda_J = 111.1$, $1/\lambda_A = 2500$, $K = 449$ for common frogs (see Materials and Methods for further parameter definitions and model details). In the modelled population, the transient increase of migrating adults is achieved as follows. The population is stable before the installation of tunnels. After the tunnels have been installed, the overall success of the reproduction is increased leading to an augmented juvenile population that leads to the increase of migrating adults with a four year delay after the installation of the road mitigation system. Consequently, the model predicts that the population of juveniles is significantly increased as early as 1992 (Figure 3b). The model therefore successfully explains why the population of adults is affected with a four year delay. This is the time needed for the excess juveniles hatched in 1992 to reach the adult stage. It also provides an explanation for the short-lasting effect in adult population. A mild increase in adult population saturates the carrying capacity of the environment. Consecutively, competition for resources is increased exerting a negative effect on juveniles and adult classes. On the long-term, the adult population returns to a level very similar to the level observed before tunnel installation (compare levels in 1983-1992 and 2000-2004 as predicted by the model). On the other hand, the juvenile population is significantly and durably increased (Figure 3b).

# DISCUSSION AND CONCLUSION

Massive amphibian road kills during seasonal migrations have attracted a lot of attention because of their striking aspect [7-11,16-19]. In this paper, we reported the results of a long-term survey of two amphibian populations at a single site in Switzerland where permanent under-road tunnels were installed in migration corridors as a road mitigation measure. We observed that the adult populations censused at this location do not exhibit any long-term trend following the installation of the tunnels but exhibit a very significant transient increase four years after their installation. In a second step, we have shown that this specific pattern cannot be explained by any other data at our disposal. More specifically, we could rule out specie-specific effects, the level of the lake that floods the breeding sites, the ground topography, management measures, impact of climate on reproduction, predation and finally, possible contributions from a distant group of population. Of course, this does not demonstrate that the specific pattern observed in this area was indeed induced by the road mitigation measure. Repetitions of such a survey will be needed to confirm or infirm this hypothesis. However, this constitutes a significant enough body of evidence to consider the contribution of the road mitigation measure as a plausible cause to the observed migration pattern. We then showed using a simple stage-class population model that the road mitigation measure together with competition for resources can successfully explain the experimental observations. In the following, we would like to discuss what the implications for conservation planners are if our model was to be supported by repetitions of our experimental observations.

Our model study suggests that the increase in adult migrating populations is explained by the installation of the road mitigation measure. The subsequent decrease in adult migrating populations is then explained by the increased competition for resources. A dense transportation network with high traffic load is one of the symptoms of important human land use [13]. Road kills are therefore likely to happen in an environment where amphibian habitat has been already reduced in significant proportions. Our model study therefore suggests that both these aspects need to be considered when planning the installation of road mitigation measures. Any road mitigation measure in a highly urbanized area like the Swiss plateau should then produce the type of pattern that we report here (delayed and short-lasting peak in the migrating population; Figure 1a). We observe that the migrating population was not significantly affected by the tunnels in years immediately following their installation (1993-1995) suggesting that the individuals killed on the road were being replaced by new adults at an equivalent rate. Taken together with the fact that no long-term sustained positive effect was observed in the migrating populations, these findings seriously question the efficacy of tunnels in terms of restoring adult breeding populations. While the tunnels do prevent road kills efficiently, our model study suggests that their installation in the Cheseaux area may have intervened too late after significant habitat reduction. These results illustrate the importance of preserving enough appropriate habitats for amphibians [15] and wildlife in general [see e.g. the recent controversy about the Florida panther (*Puma concolor coryi*) [37]]. In the specific case of amphibians, it also suggests that efficient conservation measures cannot be designed in terms of pond management only. The potential size and quality of wintering sites should be taken into account when deciding where to implement amphibian road mitigation measures. Alternatively, appropriate forest conservation measures should be applied conjointly to the installation of under-road tunnels.

We see a completely different picture when considering the modelled impact of the road mitigation measure on juveniles. Although the adult migrating population is only mildly affected on the long-term by the installation of tunnels, our model study suggests in agreement with partial censuses of returning juveniles that the overall population is strongly reinforced by a build up of the juvenile classes. This would clearly be an important positive effect since these additional juveniles would limit the risk of local stochastic extinction. Furthermore, always in our model, many more juveniles are hatched that leave the system. This is true as well for most of the adults contributing to the peak migration of 1996 which did not participated to the spring migration in consecutive years. While the model assumes for simplicity that these individuals die because of competition for resources, they could as well emigrate (competition-driven migration) from the Grande Cariçaie natural reserve to neighbouring ponds in the countryside, maybe after long-range migrations [4,32,38-40]. Although this needs further assessment, this hypothesis is partially confirmed by the fact that no massive mortality event was ever reported in years 1996-2004 in the reserve. The newly created reservoir could be important to diminish inbreeding depression and genetic drift in isolated populations from nearby urban areas [41,42] and to restore populations after local extinctions or in recreated wetlands [32,43]. In this context, extensive studies of populations at regional scale are needed if we want to understand spatial dynamics and more specifically, whether amphibians are as philopatric as it has been suggested in the past or not [4,32,33,39,44].

In regard of our results, it is urgent to replicate similar surveys of amphibian populations at the sites where road mitigation measures will be installed. As illustrated by our model study, particular attention should be put on migrating juvenile populations to assess whether or not the juvenile classes are affected by the conservation measure. If our hypothesis and model were to be confirmed, amphibian tunnels should then be considered as a very potent conservation and mitigation measure in highly urbanized areas even though their installation may not produce striking long-term increase of the censused populations. They could help consolidate local populations and very likely the regional metapopulation as well.

## *Acknowledgments*

The authors would like to thank the Service des routes de l'Etat de Vaud for providing free access to part of the data presented in this paper.

## References

1. Alford RA, Richards SJ (1999) Global amphibian declines: A problem in applied ecology. Ann Rev Ecol Syst 30: 133-165.
2. Pounds JA, Fogden MPL, Campbell JH (1999) Biological response to climate change on a tropical mountain. Nature 398: 611-615.
3. Houlahan JE, Findlay CS, Schmidt BR, Meyer AH, Kuzmin SL (2000) Quantitative evidence for global amphibian population declines. Nature 404: 752-755.
4. Green DM (2003) The ecology of extinction: population fluctuation and decline in amphibians. Biol Conserv 111: 331-343.
5. Berger L, Speare R, Daszak P, Green DE, Cunningham AA, et al. (1998) Chytridiomycosis causes amphibian mortality associated with population declines in the rain forests of Australia and Central America. Proc Natl Acad Sci USA 95: 9031-9036.
6. Kiesecker JM, Blaustein AR, Belden LK (2001) Complex causes of amphibian population declines. Nature 410: 681-684.
7. Fahrig L, Pedlar JH, Pope SE, Taylor PD, Wegner JF (1995) Effect of road traffic on amphibian density. Biol Conserv 73.
8. Vos CC, Chardon JP (1998) Effects of habitat fragmentation and road density on the distribution pattern of the moor frog *Rana arvalis*. J Appl Ecol 35: 44-56.
9. Carr LW, Fahrig L (2001) Effect of Road Traffic on Two Amphibian Species of Differing Vagility. Conserv Biol 15: 1071-1078.
10. Pellet J, Guisan A, Perrin N (2004) A concentric analysis of the impact of urbanization on the threatened European tree frog in an agricultural landscape. Conserv Biol 18: 1599-1606.
11. Gibbs JP, Shriver GW (2005) Can road mortality limit populations of pool-breeding amphibians? Wetl Ecol Manag 13: 281-289.
12. Spellerberg IF (1998) Ecological effects of roads and traffic: a literature review. Global Ecol Biogeogr Lett 7: 317-333.
13. Foley JA, DeFries R, Asner GP, Barford C, Bonan G, et al. (2005) Global Consequences of Land Use. Science 309: 570-574.
14. Mader H-J (1984) Animal habitat isolation by roads and agricultural fields. Biol Conserv 29: 81-96.
15. Johnson B (1992) Habitat loss and declining amphibian populations. In: Bishop CA, Pettit KE, editors. Declines in Canadian Amphibian Populations: Designing a National Monitoring Strategy. Ottawa: Occas Pap Can Wildlife Serv. pp. 71-75.
16. van Gelder JJ (1973) A quantitative approach to the mortality resulting from traffic in a population of *Bufo bufo* L. Oecologia 13: 93-95.
17. Cooke AS (1995) Road mortality of common toads (*Bufo bufo*) near a breeding site. Amphibia-Reptilia 16: 87-90.
18. Ashley EP, Robinson JT (1996) Road mortality of amphibians, reptiles and other wildlife on the Long Point Causeway, Lake Erie, Ontario. Can Field-Nat 110: 403-412.
19. Hels T, Buchwald E (2001) The effect of road kills on amphibian populations. Biol Conserv 99: 331-340.
20. Langton TES, editor (1989) Amphibians and Roads: Proceedings of the Toad Tunnel Conference. Shefford: ACO Polymer Products. 202 p.
21. Ceballos G, Ehrlich PR, Soberón J, Salazar I, Fay JP (2005) Global Mammal Conservation: What Must We Manage? Science 309: 603-607.
22. Office fédéral de la statistique (2001) L'utilisation du sol: hier et aujourd'hui. Neuchâtel: Office fédéral de la statistique. 32 p.
23. Ryser J, Grossenbacher K (1989) A survey of amphibian preservation at roads in Switzerland. In: Langton TES, editor. Amphibians and Roads: Proceedings of the Toad Tunnel Conference. Shefford: ACO Polymer Products. pp. 7-13.
24. Günther R, editor (1996) Die Amphibien und Reptilien Deutschlands. Jena: G. Fischer. 825 p.
25. Commission technique VSS2 (2001) Swiss Standard SN 640 005a. Zürich: Association suisse des professionnels de la route et du trafic. 44 p.
26. Lewis EG (1942) On the generation and growth of a population. Sankhya 6: 93-96.
27. Leslie PH (1945) On the use of matrices in certain population mathematics. Biometrika 33: 183-212.
28. Lefkovitch LP (1965) The study of population growth in organisms grouped by stages. Biometrics 21: 1-18.

29. Nelder J, Mead R (1965) A simplex method for function minimization. Comput J 7: 308-313.
30. Meyer AH, Schmidt BR, Grossenbacher K (1998) Analysis of three amphibian populations with quarter-century long time-series. Proc R Soc Lond B 265: 523-528.
31. Hanski I (1998) Metapopulation dynamics. Nature 396: 41-49.
32. Marsh DM, Trenham PC (2001) Metapopulation Dynamics and Amphibian Conservation. Conserv Biol 15: 40-49.
33. Sinsch U (1990) Migration and orientation in anuran amphibians. Ethol Ecol Evol 2: 65-79.
34. Hoppensteadt FC, Keller JB (1976) Synchronization of periodical cicada emergences. Science 194: 335-337.
35. Miaud C, Guyétant R, Elmberg J (1999) Variations in life-history traits in the common frog *Rana temporaria* (Amphibia: Anura): a literature review and new data from the French Alps. J Zool 249: 61-73.
36. Drake JM (2005) Density-Dependent Demographic Variation Determines Extinction Rate of Experimental Populations. PLoS Biol 3: e222.
37. Gross L (2005) Why Not the Best? How Science Failed the Florida Panther. PLoS Biol 3: e333.
38. Loman J (1981) Spacing mechanisms in a population of the common frog *Rana temporaria* during the non-breeding period. Oikos 37: 225-227.
39. Schlupp I, Podloucky R (1994) Changes in breeding site fidelity: a combined study of conservation and behavior in the common toad *Bufo bufo*. Biol Conserv 69: 285-291.
40. Sinsch U (1997) Postmetamorphic dispersal and recruitment of first breeders in a *Bufo calamita* metapopulation. Oecologia 112: 42-47.
41. Sjögren-Gulve P (1994) Distribution and extinction patterns within a northern metapopulation of the pool frog, *Rana lessonae*. Ecology 75: 1357-1367.
42. Hitchings SP, Beebee TJC (1997) Genetic substructuring as a result of barriers to gene flow in urban *Rana temporaria* (common frog) populations: implications for biodiversity conservation. Heredity 79: 117-127.
43. Lehtinen RM, Galatowitsch SM (2001) Colonization of Restored Wetlands by Amphibians in Minnesota. Am Midl Nat 145: 388-396.
44. Blaustein AR, Wake DB, Sousa WP (1994) Amphibian Declines: Judging Stability, Persistence, and Susceptibility of Populations to Local and Global Extinctions. Conserv Biol 8: 60-71.

# SUPPLEMENTARY FIGURES

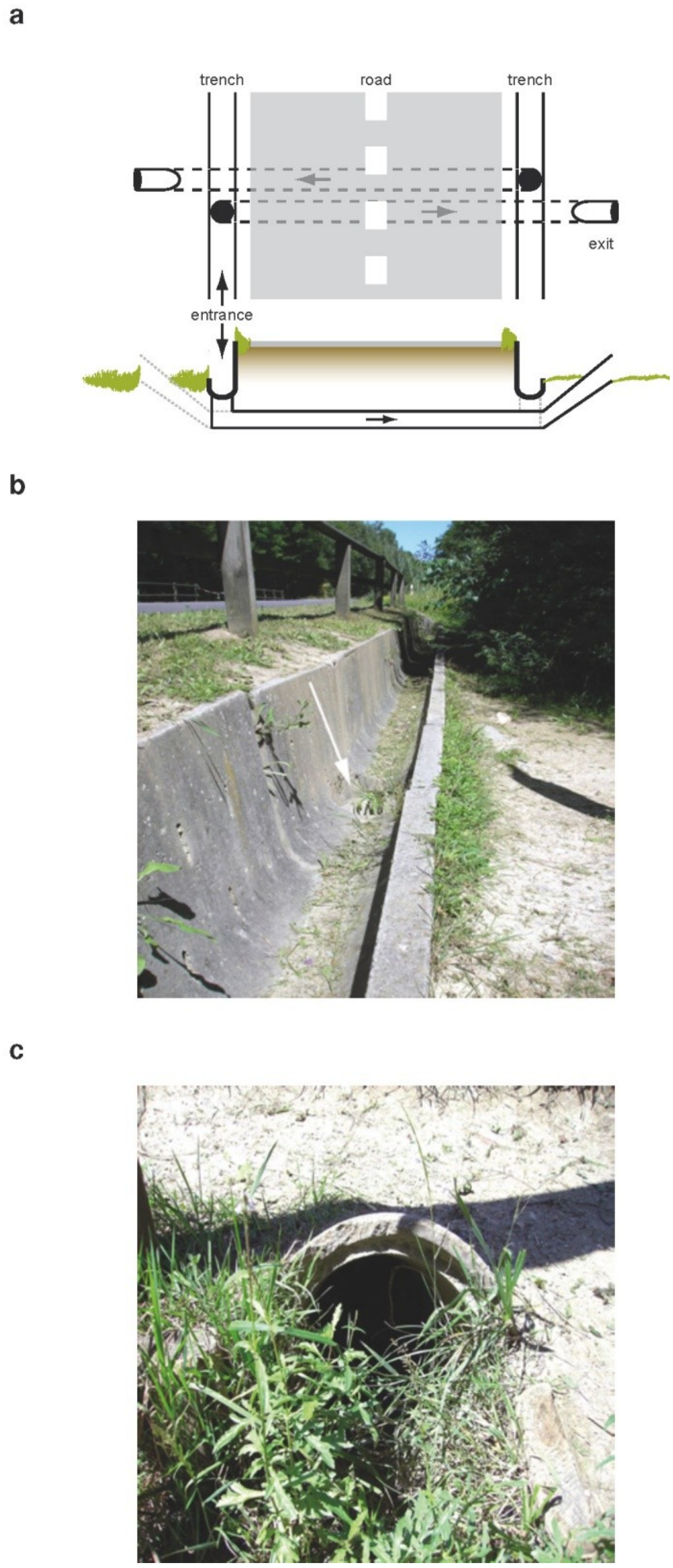

FIGURE S 1 DETAILS OF THE AMPHIBIAN TUNNELS. (A) DIAGRAM OF THE INSTALLATION. ON EACH SIDE OF THE ROAD, A CONCRETE TRENCH WAS CONSTRUCTED THAT PREVENTS AMPHIBIANS FROM PASSING ONTO THE ROAD. (B) FROM THE TRENCH, AMPHIBIANS FALL IN A TUNNEL (WHITE ARROW) WHOSE ONLY EXIT. (C) IS ON THE OTHER SIDE OF THE ROAD. THE TUNNEL PASSES BOTH UNDER THE ROAD AND UNDER THE OPPOSITE TRENCH.

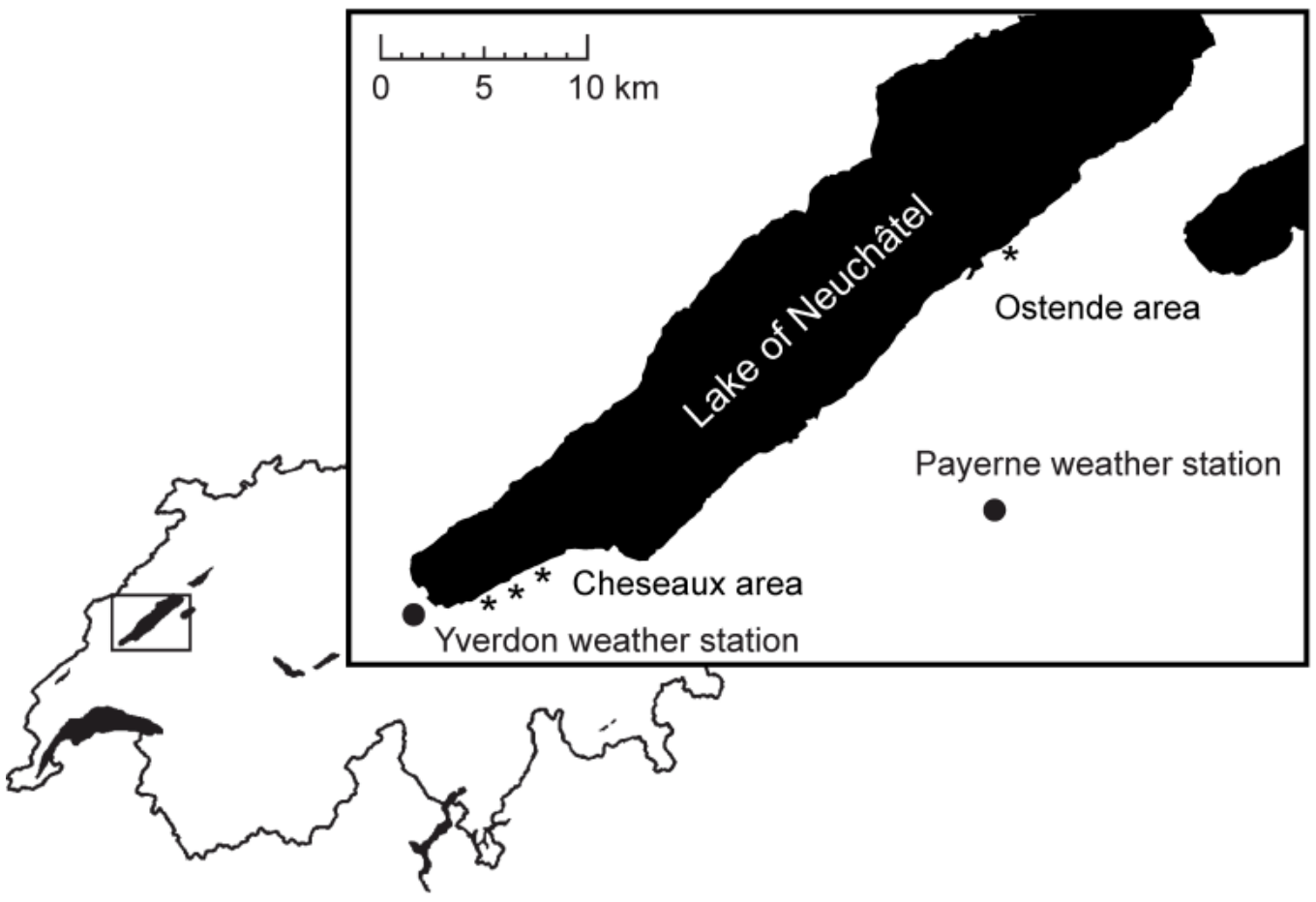

FIGURE S 2 SITUATION MAP. STARS DENOTE THE LOCATION OF THE THREE SECTORS PROTECTED WITH AMPHIBIAN TUNNELS (CHESEAUX) AND OF THE DRIFT FENCES (OSTENDE). THE WEATHER STATION IN PAYERNE IS THE STATION CLOSEST TO THE OSTENDE AREA ON THE EASTERN SIDE OF THE LAKE.

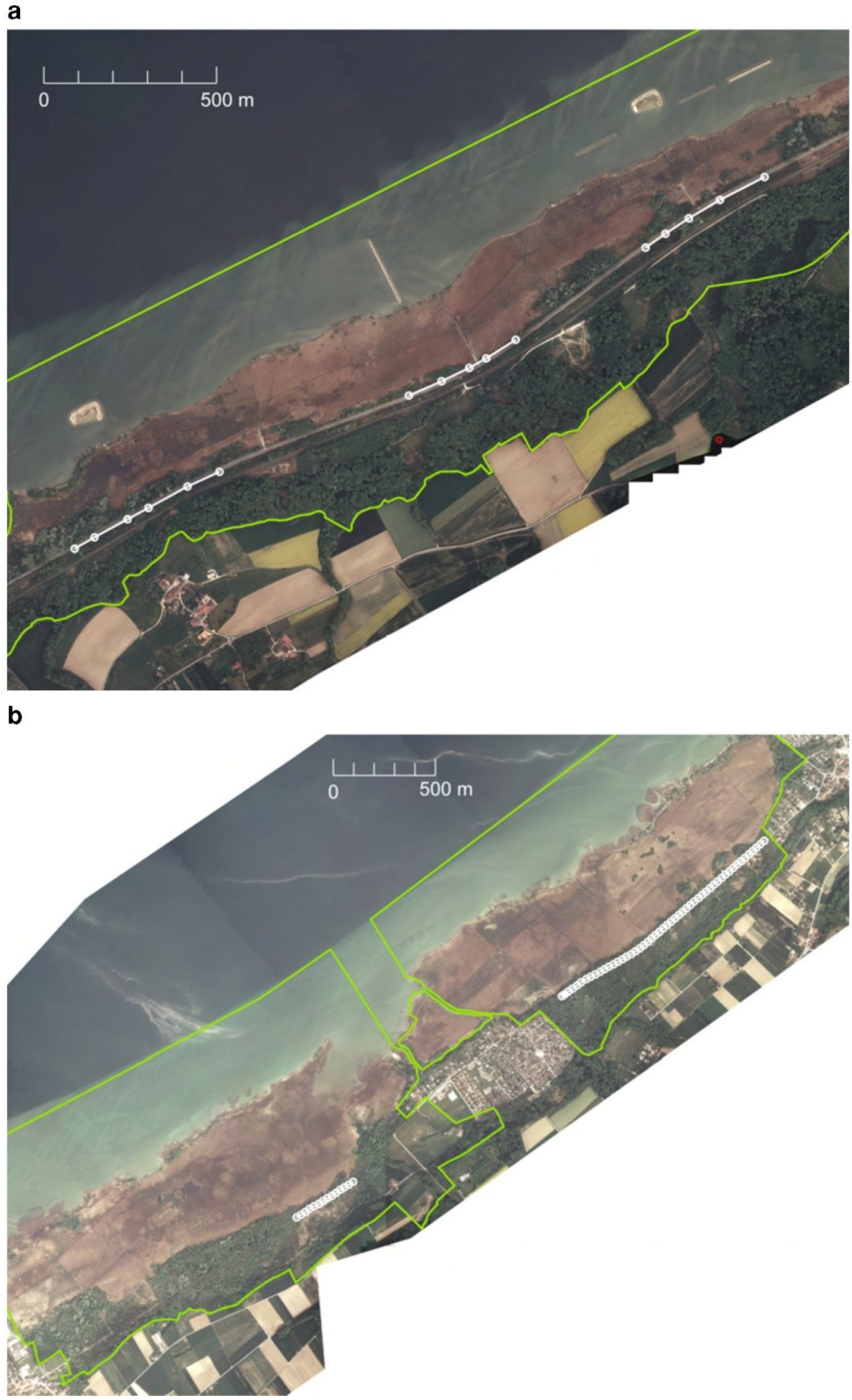

FIGURE S 3 AERIAL VIEWS OF CHESEAUX (A) AND OSTENDE AREAS (B). THE GREEN LINES MARK THE BORDER OF THE NATURAL RESERVE. IN BOTH AREAS, CENSUSES WERE PERFORMED ALONG THE MAIN MIGRATION CORRIDORS. THE WHITE LINES MARK THE POSITION OF PROTECTED SECTORS (A) AND DRIFT FENCES (B) WITH CAPTURE POINTS (CIRCLES; BOW-NETS OR PIT TRAPS, SEE MATERIALS AND METHODS).

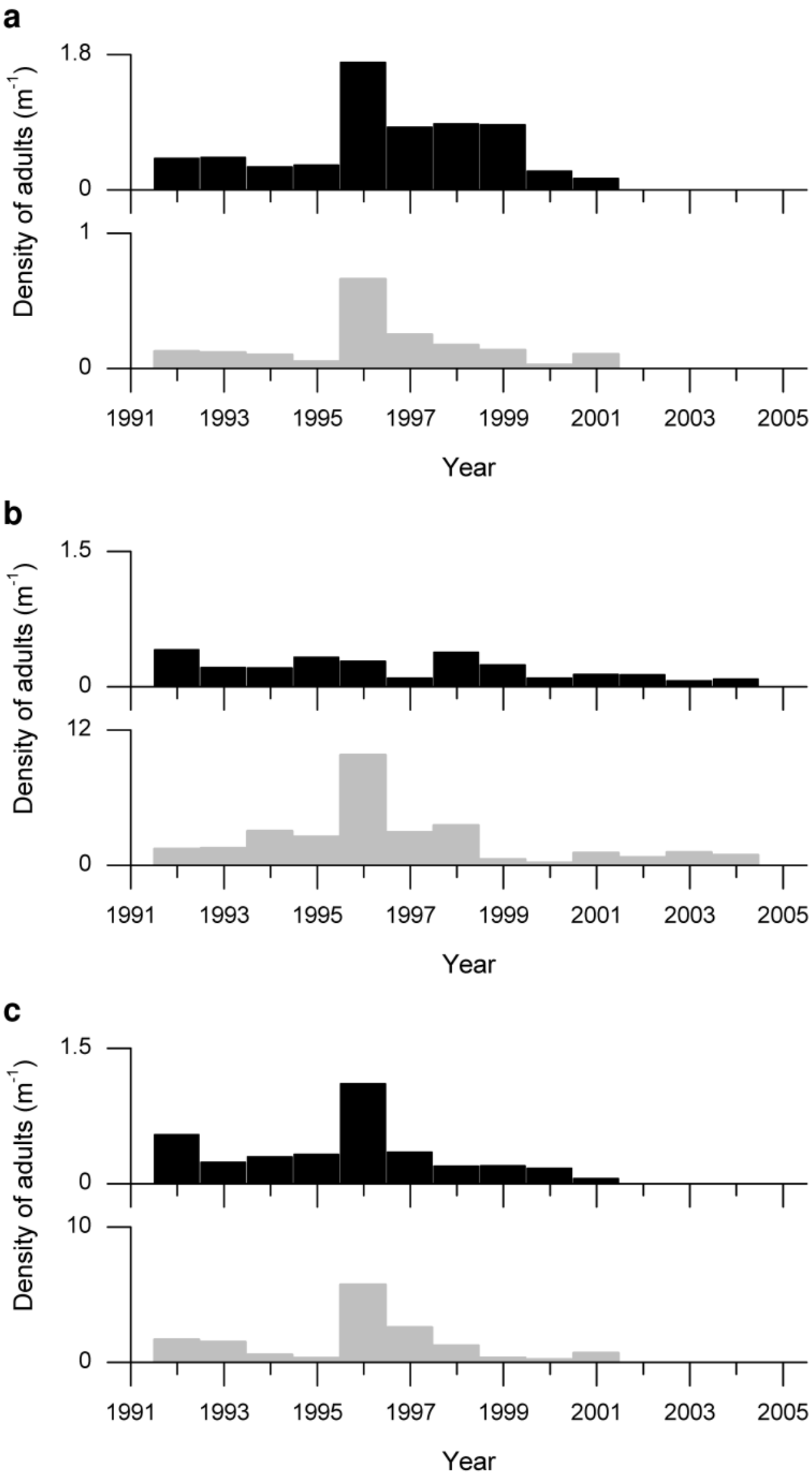

FIGURE S 4 DETAILS OF THE CENSUS IN THE CHESEAUX AREA. DENSITIES OF COMMON TOADS (*BUFO BUFO*; BLACK BARS; NUMBER OF MIGRATING ADULTS CAPTURED PER METER OF TRENCH) AND COMMON FROGS (*RANA TEMPORARIA*; GREY BARS) CENSUSED IN THE SOUTHERN TRENCH (A), IN THE CENTRAL TRENCH (B) AND IN THE NORTHERN TRENCH (C).